# Directly diode-pumped, Kerr-lens mode-locked, few-cycle Cr:ZnSe oscillator

NATHALIE NAGL,[1,3] SEBASTIAN GRÖBMEYER,[1] VLADIMIR PERVAK,[1] FERENC KRAUSZ,[1,2] OLEG PRONIN,[2] AND KA FAI MAK[2,4]

[1]Ludwig-Maximilians-Universität München, Am Coulombwall 1, 85748 Garching, Germany
[2]Max-Planck Institute of Quantum Optics, Hans-Kopfermann-Str. 1, 85748 Garching, Germany
[3]e-mail: nathalie.nagl@physik.uni-muenchen.de
[4]e-mail: kafai.mak@mpq.mpg.de

Lasers based on $Cr^{2+}$-doped II-VI material, often known as the Ti:Sapphire of the mid-infrared, can directly provide few-cycle pulses with super-octave-spanning spectra, and serve as efficient drivers for generating broadband mid-infrared radiation. It is expected that the wider adoption of this technology benefits from more compact and cost-effective embodiments. Here, we report the first directly diode-pumped, Kerr-lens mode-locked $Cr^{2+}$-doped II-VI oscillator pumped by a single InP diode, providing average powers of over 500 mW and pulse durations of 45 fs — shorter than six optical cycles at 2.4 µm. These correspond to a sixty-fold increase in peak power compared to the previous diode-pumped record, and are at similar levels with respect to more mature fiber-pumped oscillators. The diode-pumped femtosecond oscillator presented here constitutes a key step towards a more accessible alternative to synchrotron-like infrared radiation, and is expected to accelerate research in laser spectroscopy and ultrafast infrared optics.

## 1. INTRODUCTION

The mid-infrared (mid-IR) spectral region between 500 $cm^{-1}$ and 4000 $cm^{-1}$ (2.5-20 µm) contains an immensely rich set of unique molecular absorption lines, and is thus often known as the molecular 'finger-print' region. Consequently, broadband light sources covering this range enable species-specific, sensitive, and quantitative measurements of organic compounds and have been widely applied in chemical and biological analysis [1], such as trace gas sensing [2] and early cancer detection [3]. The spatial resolution of such investigations can be markedly improved by using radiation with high brightness [3], for example those from synchrotron facilities [4], with which multidimensional infrared absorption images can be obtained for nanoscale samples [5-8]. Recently, similar broadband high-brightness mid-infrared radiation have also been generated using high-power femtosecond lasers [9-11] at outstanding wavelength and power stability [12]. More intriguingly, the femtosecond nature of such output, especially when in the form of a stabilized frequency comb, opens the way to a multitude of time- and frequency-domain techniques [13-15] that can reveal ultrafast dynamics and drastically improve the speed, dynamic range, and many other aspects of spectroscopic measurements [16-20].

Solid-state lasers based on $Cr^{2+}$-doped II-VI material, often called the Ti:Sapphire of the mid-infrared, are reliable sources for directly generating ultrashort femtosecond pulses in the 2-3 µm spectral range [21-25]. With recent improvements in output power to Watt-level and pulse durations down to few optical cycles [26, 27], the output has also been effectively down-converted to generate broadband mid-infrared radiation between 3–20 µm [28-30].

To date, virtually all mode-locked (ML) Cr:ZnS/ZnSe lasers are pumped by mature rare-earth-doped fiber lasers. Thanks to the near diffraction-limited fiber output and high average powers, Kerr-lens mode-locked (KLM) oscillators with up to 2 W of average power have been demonstrated [31]. On the other hand, direct diode-pumping represents an excellent alternative that combines compactness, high-efficiency, affordability, and low-noise performance. Extending it to femtosecond Cr:ZnS/ZnSe lasers will help meet the growing demand for reliable and accessible table-top mid-IR sources [32]. Although Cr:ZnS/ZnSe crystals can be directly pumped at the center of their absorption band using laser diodes emitting at around 1.5–1.6 µm, the optical output powers of these diodes have been relatively low compared to diodes operating at shorter wavelengths, mainly due to temperature-related problems [33, 34]. Continuous-wave (CW) diode-pumped Cr:ZnS/ZnSe oscillators were first demonstrated in a side-pumped configuration [35], followed by longitudinal pumping [36]. Since then, output powers of over 400 mW were reported when using single-emitter diode lasers [37], while even higher output power of up to several Watts has been demonstrated when the oscillator is pumped with a 30 W diode bar [38]. Though diode bars can provide high powers, the beam quality is, however, very much reduced. This leads to a shorter Rayleigh length at tight foci and a reduced overlap, in longer gain crystals, between the pump beam and the fundamental laser mode. The resulting reduction in conversion efficiencies and increased heat load makes diode bars less suitable for pumping Cr-doped II-VI, which is sensitive to thermal distortion [39]. For femtosecond operation, and especially for Kerr-lens mode-locking of bulk-crystal oscillators, where operation in the fundamental spatial mode is desirable, utilizing the higher brightness available from a single-emitter diode is thus important [40].

The first femtosecond operation of a $Cr^{2+}$-doped II-VI oscillator directly pumped by a 1550 nm, 3.5 W single-emitter laser diode provided an average output power of approximately 50 mW [41]. Although the use of semiconductor saturable absorber mirrors (SESAM) allowed mode-locking to self-start, the resultant pulse duration of 180 fs was significantly longer than allowed by the gain medium. On the contrary, Kerr-lens mode-locking is known to support much shorter pulse durations [42], even beyond the emission bandwidth-limit of the laser gain medium [43, 44].

In this work, we report the first directly diode-pumped, Kerr-lens mode-locked $Cr^{2+}$-doped II-VI oscillator, generating pulses at a repetition rate of 64.7 MHz and pulse durations of 45 fs — comprising less than six optical cycles at 2.4 µm. With average output powers of over 500 mW, this represents a ten-fold increase in average power as well as a four-fold reduction in pulse duration compared to the previous demonstration [41]. It also constitutes an over 2.5-times increase in peak power over more mature fiber-pumped Cr:ZnSe oscillators [24, 45]. In addition, amplitude noise measurements of the frequency-doubled mode-locked pulse train have revealed remarkable low-noise performance (integrated root-mean-square (RMS) relative intensity noise (RIN) < 0.068 % (20 Hz - 1 MHz)).

## 2. RESULTS AND DISCUSSION

### A. Diode-pumped KLM Cr:ZnSe oscillator

The experimental setup of the Kerr-lens mode-locked oscillator is depicted in Fig. 1(a) and is based on a 5 mm-long polycrystalline Cr:ZnSe gain element (*IPG Photonics*). The anti-reflection (AR) coated crystal is placed inside an asymmetric X-fold cavity at normal-incidence angle [31, 41], and pumped by a single InP C-Mount laser diode at 1650 nm (*SemiNex Corp.*) providing up to 3.6 W from a 95 µm broad stripe. Furthermore, thermoelectric cooling was implemented to stabilize the diode's output power and wavelength. Using a scanning slit profiler (*Ophir*, NanoScan 2s), the $M^2$-values of the fast (vertical) and slow (horizontal) axis were measured to be 1.2 and 7.6, respectively. The highly-divergent and anisotropic diode output was therefore reshaped with a set of AR-coated aspheric and cylindrical lenses, analogous to diode-pumped KLM Ti:Sapphire laser systems [40, 46, 47]. The diode beam was then focused through a 6 mm-thick input coupler (IC) mirror placed at 45 degrees. A focal spot diameter of approximately 50 µm (fast axis) x 90 µm (slow axis) at the gain crystal position was experimentally measured, being in good agreement with the results obtained by ABCD-matrix calculation using the open source software *reZonator* (Fig. 1(b)).

The AR-coated Cr:ZnSe crystal was mounted on top of a water-cooled copper heat sink cooled to 12°C, with the measured single-pass absorption amounting to ~90 % at the 1.65 µm pump wavelength. It was placed between two curved mirrors (CM1, CM2), which have different radii of curvature (ROC) due to the limited selection of suitable optics at the time of the oscillator construction. To minimize astigmatism inside the cavity, the angle of incidence on the spherical mirrors was kept small at around 8 degrees, and further compensated by a 3 mm-thick YAG plate (M1) placed in the focusing beam between the crystal and one of the curved mirrors. The use of multiple highly-reflective 45-degree mirrors allowed the entire oscillator, excluding the pump setup, to be assembled inside a monolithic housing with a footprint of only 30 cm x 70 cm.

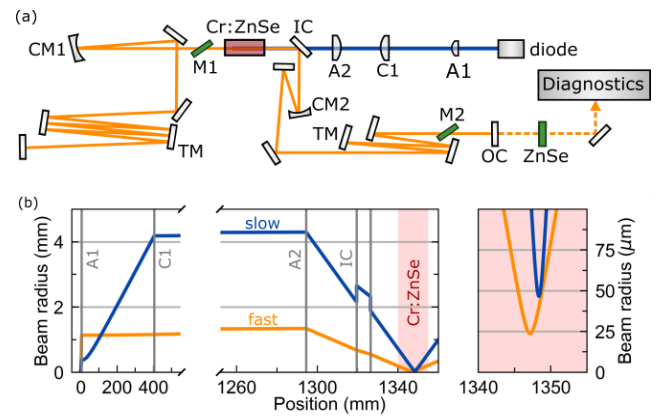

**Fig. 1.** Schematic of the setup and simulated pump propagation. (a) Experimental setup. CM: curved mirror (CM1: ROC=200 mm, CM2: ROC=100 mm), A: aspheric lens (A1: f=4.51 mm, A2: f=50 mm), C1: convex cylindrical lens (f=400 mm), M: Brewster-oriented material (M1: 3 mm YAG, M2: 3 mm sapphire), TM: TOD mirror, OC: output coupler (34 %), ZnSe: 5 mm AR-coated substrate used to re-compress the pulses. Non-labelled mirrors as well as the input coupler (IC) are highly-reflective coated (2.0 µm-2.6 µm) and transmit the pump light. (b) Simulated pump propagation for the fast and slow axis. The theoretical pump spot diameter of 48 µm (fast) x 94 µm (slow), calculated from the measured beam divergences and radii, matches very well with the experimental results (50 µm x 90 µm).

Soft-aperture Kerr-lens mode-locking was initiated by rapidly moving one end-mirror. The oscillator provided over 700 mW of CW and 500 mW of ML average power with RMS power fluctuations of less than 0.12 % over 3 hours (Fig. 2(e)). To compensate the cavity dispersion and generate ultrashort pulses, a 3 mm-thick plane-parallel sapphire plate (M2) was introduced at Brewster angle to balance the group-delay dispersion (GDD) of the gain crystal and M1. Additional mirrors manufactured in-house compensated for the cavity's third-order dispersion (TOD) and helped to shorten the pulse duration further (see the supplementary material). With a self-built second-harmonic frequency-resolved optical gating device (SHG-FROG), the full width at half maximum (FWHM) pulse duration was determined after compensating for the dispersion of the 6.35 mm-thick fused-silica output coupler (OC) substrate (Fig. 2(a)-(d)). We measured pulses as short as 45 fs (error: 0.7 x 10$^{-3}$, grid size: 512 x 512) at a repetition rate of 64.7 MHz and an output coupling ratio of 34 %, resulting in a peak power of 159.4 kW. The pulse duration is close to the Fourier-transform limit of 44 fs and corresponds to less than six optical cycles at 2.4 μm. In addition, using a scanning slit profiler, the horizontal and vertical beam quality factor M$^2$ of the ML output pulse trains were characterized to be nearly diffraction limited with M$^2$<1.1 (Fig. 2(e), Inset), with the beam being polarized in the horizontal direction with a polarization extinction ratio of ~300.

Compared to the first femtosecond diode-pumped Cr:ZnSe oscillator based on SESAM mode-locking [41], we demonstrate a sixty-fold increase in peak power as well as an over 2.5-times increase in peak power over previously demonstrated fiber-pumped Cr:ZnSe lasers [24, 45]. Currently, the maximum output power of our oscillator is not limited by the available pump power. Excessive heating of the gain medium, when more than 2.7 W of pump power is absorbed, seems to be the obstacle to further increase in output power. Although Cr:ZnSe shares similar spectroscopic characteristics with Cr:ZnS, it has an about 1.5 times higher thermo-optic coefficient [48]. Furthermore, since the crystal is currently cooled mainly from one side only, an improved cooling architecture in the future can potentially allow the combining of two or more separate pump diodes, boosting the average output power to the Watt-level.

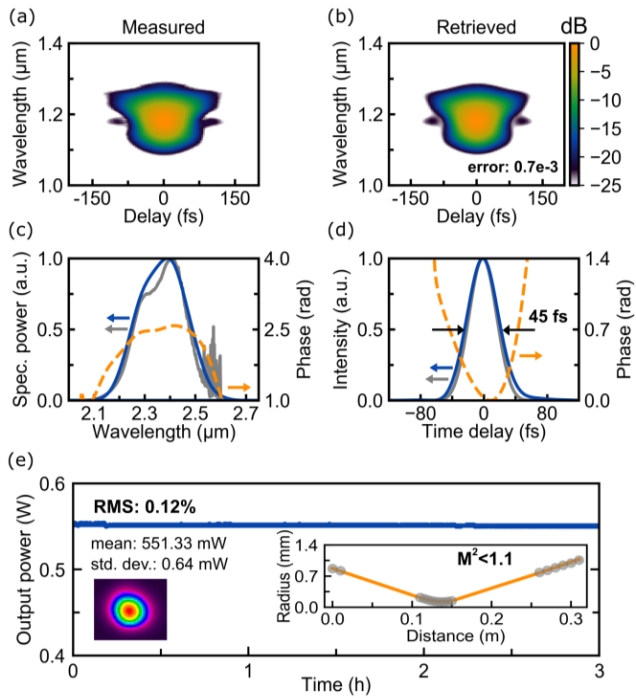

**Fig. 2.** Output characteristics of the oscillator. (a)-(b) Measured and retrieved FROG traces for temporal characterization. (c) Retrieved FROG results in the spectral domain, including the measured output spectrum in grey for comparison. (d) Temporal profile of the retrieved pulse, showing a FWHM pulse duration of 45 fs and the corresponding Fourier-transform limit of 44 fs in grey-scale. (e) Mode-locked output power stability over 3 hours, where the inset shows the corresponding beam profile and the M$^2$-measurement for the horizontal/vertical axis.

## B. Amplitude noise measurements

Applications of femtosecond mid-infrared radiation such as FTIR-microscopy or s-SNOM can highly benefit from compact and affordable high-brightness light sources. However, they also require low-noise laser performance since intensity fluctuations will set a limit to the sensitivity of measurements. Therefore, the amplitude noise characteristics of our mode-locked pulse trains were carefully characterized, using a high dynamic-range radio-frequency (RF) spectrum analyzer in combination with a biased InGaAs photodetector. Applying the procedure described in the supplementary material, the noise of the direct oscillator output was measured first. Surprisingly, the amplitude noise was already close to the noise floor of the measurement devices when the photodetector was operating at the limit of the linear regime. Therefore, additional data was taken on the frequency-doubled output generated in a PPLN crystal (Fig. 3, blue curves), which should enhance the relative intensity noise by a factor of two [49] given the quadratic dependence of the frequency conversion process on the optical input power. This corresponds to an increase of 6 dB in the electrical power. To block contributions from the fundamental and other wavelengths, a bandpass filter centered at 1.2 µm was placed in front of the detection setup (see the supplementary material). Two consecutive measurements were performed with different resolution bandwidths (RBWs) and frequency spans, where the measurement noise floor was obtained by blocking the optical signal to the detector (Fig. 3, grey curves).

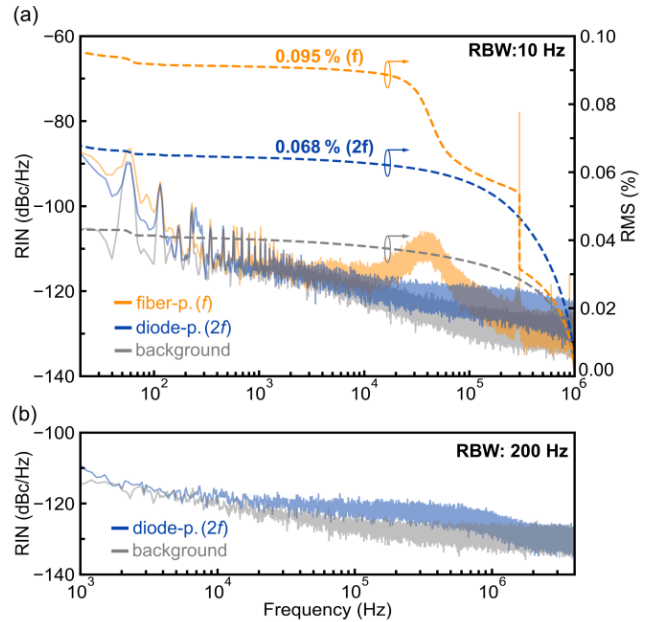

**Fig. 3.** Amplitude noise measurements. (a) The RIN and integrated noise are shown for the frequency-doubled (2$f$) diode-pumped laser pulses (blue) and the background (grey). The data was taken in a 1 MHz-span, with the integration starting at 1 MHz and ending at 20 Hz. For comparison, the corresponding data of the fundamental ($f$) is added when the oscillator was pumped by a commercial fiber laser. (b) The relative intensity noise was measured in a broader range up to 4 MHz, where the noise signal drops back to the background noise floor at frequencies around 2 MHz.

From Fig. 3 (blue curves), it can be seen that the noise spectrum exhibits a $f^{-1}$-behaviour [50] at the lower frequencies up to 200 Hz. Beyond 1 kHz, it is relatively smooth, similar to white noise, and falls to the measurement noise floor near 2 MHz. It is known that pump noise can be transferred to oscillator noise below the oscillator's relaxation oscillation frequency [51]. Given the short fluorescence lifetime of Cr:ZnSe, which is on the order of a few microseconds [52], the filtering effect of the gain medium is likely to cause the drop in noise at higher frequencies. It is worth noting that the usual noise peak originating from relaxation oscillations in erbium and thulium fiber lasers at hundreds of kHz [24] is absent in this directly diode-pumped setup. Particularly, the RIN measured in Fig. 3 shows no clear noise peak from the Cr:ZnSe output. Instead, a more 'plateau'-like noise structure is seen. Similar behaviour has also been observed for diode-pumped Ti:Sapphire oscillators, which has been attributed to the relaxation oscillations of the Ti:Sapphire laser itself [53, 54].

To further investigate the transfer of pump noise, the pump source of the Cr:ZnSe oscillator was switched to a commercial fiber laser (*Bktel Photonics*, HPFL-370-1567). This fiber laser was not optimized for low-optical-noise performance, and is not a representation of state-of-the-art low-noise fiber lasers. Nevertheless, it serves to illustrate the noise characteristics of the Cr:ZnSe oscillator when pumped with a noisy source. In contrast to the noise analysis shown for the frequency-doubled diode-pumped oscillator output, the fluctuation in optical power of the fiber-pumped oscillator was measured at the fundamental wavelength, i.e. without further noise-enhancing frequency doubling. As shown in Fig. 3(a) (orange curve), the resulting optical noise of the fiber-pumped Cr:ZnSe output is notably higher, with spectral features originating from the fiber laser being clearly transferred through the system, as also observed by Wang et al. [24].

To provide a quantitative comparison between the diode- and fiber-pumped Cr:ZnSe oscillator, the RIN is integrated over the measured frequency span from 20 Hz to 1 MHz. As the RF analyzer itself exhibited undesirably high noise below 20 Hz, even when simply terminated with a 50 Ω resistor, the lower integration limit was set to 20 Hz. This resulted in RMS noise levels of 0.095 % for the fiber-pumped Cr:ZnSe oscillator, and 0.043 % for the background of the measurement setup. In comparison, for the directly diode-pumped Cr:ZnSe oscillator, the integrated RIN of the frequency-doubled output is measured to be 0.068 %, and the noise of the fundamental output should be even lower. This clearly illustrates the potential of using simple, affordable, yet low-noise laser diodes for pumping Cr:ZnS/ZnSe lasers.

## 3. CONCLUSION

In summary, we demonstrate, to the best of our knowledge, the first directly diode-pumped Kerr-lens mode-locked Cr$^{2+}$-doped II-VI oscillator. Driven by a single 1.65 µm InP laser diode, the Cr:ZnSe oscillator provides an output power of over 500 mW and pulse durations as short as 45 fs — a sixty-fold increase in peak power compared to the previous work on a directly diode-pumped femtosecond Cr$^{2+}$-doped II-VI oscillator [41]. Moreover, the diode-pumped output exhibits excellent amplitude noise performance, with a measured integrated RIN of less than 0.068 % RMS (20 Hz–1 MHz) for the frequency-doubled output. By switching to a Cr:ZnS crystal and with improved crystal cooling, additional pump power from two or more diodes can be utilized to further increase the oscillator's output power [46, 47]. Careful optimization of the cavity's dispersion should also lead to even shorter pulse durations [26]. Finally, the development of diode-pumped amplifier systems can potentially boost the pulse energy further while retaining the low-noise performance [55]. Directly diode-pumped femtosecond Cr$^{2+}$-doped II-VI oscillators, which can serve as drivers for parametric down-conversion into the mid-IR region, thus represent a new class of cost-effective, low-noise, table-top sources for a broad range of spectroscopic applications.

**Funding.** This work was funded by the Munich-Centre for Advanced Photonics (MAP) and the Centre for Advanced Laser Applications (CALA). The first author received a doctoral scholarship from the Bischöfliche Studienförderung Cusanuswerk.

**Acknowledgment**. We thank Marcus Seidel for his useful insights on the noise analysis.

See supplementary material for supporting content.

# Directly diode-pumped, Kerr-lens mode-locked, few-cycle Cr:ZnSe oscillator: supplementary material

NATHALIE NAGL,[1,3] SEBASTIAN GRÖBMEYER,[1] VLADIMIR PERVAK,[1] FERENC KRAUSZ,[1,2] OLEG PRONIN,[2] AND KA FAI MAK[2,4]

[1]Ludwig-Maximilians-Universität München, Am Coulombwall 1, 85748 Garching, Germany
[2]Max-Planck Institute of Quantum Optics, Hans-Kopfermann-Str. 1, 85748 Garching, Germany
[3]e-mail: nathalie.nagl@physik.uni-muenchen.de
[4]e-mail: kafai.mak@mpq.mpg.de

**This document provides supplementary information to "Directly diode-pumped, Kerr-lens mode-locked, few-cycle Cr:ZnSe oscillator".**

## 1. INTRACAVITY DISPERSION MANAGEMENT

To exploit the broad emission bandwidth of Cr:ZnS/ZnSe gain media and obtain ultrashort pulse durations in the negative dispersion regime (soliton mode-locking), careful dispersion control is required. The tuning of the round-trip second-order dispersion was realized using a 3 mm-thick sapphire plate oriented at Brewster angle. Third-order material dispersion was compensated with mirrors providing -2800 fs$^3$ at 2.4 µm (see Fig. S1). At 2.4 µm, the round-trip GDD was approximately -300 fs$^2$, accumulated by 2 x (+1040 fs$^2$) of the 5 mm thick gain medium, 2 x (-450 fs$^2$) resulting from 3 mm YAG, and 2 x (-740 fs$^2$) due to 3 mm sapphire. However, it is important to stress that the exact dispersion values of the mirrors were not experimentally measured, and each individual dispersive mirror may show a slightly different oscillation pattern. Thus, the theoretical round-trip dispersion in Fig. S1 is only an approximation for the combination of mirrors that experimentally resulted in the shortest pulse duration.

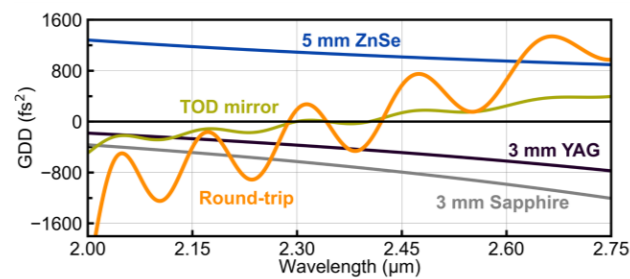

**Fig. S1.** Theoretical dispersion curves of intracavity elements. The single-pass GDD of incorporated material as well as the dispersion introduced by one TOD mirror are shown along with the total dispersion per cavity round-trip.

## 2. AMPLITUDE NOISE MEASUREMENT SETUP AND RIN CALCULATION

The layout of the noise measurement setup utilized here is depicted in Fig. S2. The pump laser diode as well as the fiber laser were driven by a low-noise current source (*Delta Elektronika*, SM 70-AR-24). A high dynamic-range radio-frequency spectrum analyzer (*Agilent*, E447A) was used together with a biased InGaAs photodetector (*Thorlabs*, DET10D2) for characterizing the noise at low frequencies between 20 Hz and 1 MHz, also referred to as baseband [1]. Due to the emission of the KLM laser being centered at around 2.4 µm, the selection of appropriate photodetectors was rather limited. With a biased InGaAs photodetector, sensitive from 0.9 µm to 2.6 µm, a signal-to-noise ratio of ~90 dB relative to the carrier-frequency could be obtained for the fundamental as well as the frequency-doubled laser pulses using a 50 Ω termination. Since beam-pointing fluctuations can cross-couple into the relative intensity noise measurements [2], the beam was focused onto the center of the detector area (Ø 1 mm) by the lens L3. To reduce mixing artefacts generated by the RF analyzer, the electronic DC signal was filtered at the analyzer input using *Thorlabs*' EF500 DC block filter. In addition, a ND filter wheel was placed in front of the photodetector to keep the optical signal below the diode saturation level.

As indicated in Fig. 3 of the manuscript, the amplitude noise was plotted in relative units of dBc Hz$^{-1}$ and enables the comparison with other systems. It was calculated by normalizing the measured RMS electrical power fluctuations to the DC voltage, which is proportional to the average optical power incident on the photodetector [3]. To achieve high signal strengths, but still operate in the linear regime of the detector, we used an average optical power of about 2.4 mW for the frequency-doubled signal and less than 1 mW for the fundamental. This corresponded to a DC voltage of 47 mV for both measurements. As can be seen in Fig. 3 of the manuscript, the background noise floor of the detection setup was located at -130 dBc Hz$^{-1}$, while the calculated shot noise level was at -154 dBc Hz$^{-1}$. Furthermore, a better estimate of the mean-squared quantities was obtained upon averaging 20 traces taken within a certain RBW, and a good overlap of the measurement points was ensured by setting the step size to 0.5 x RBW.

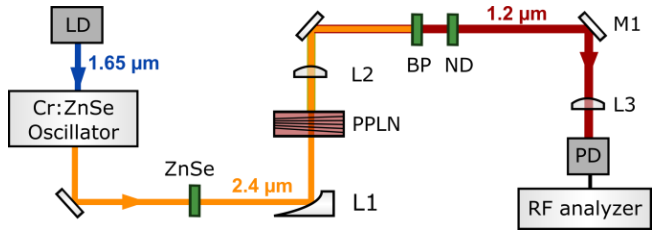

**Fig. S2.** Setup for measuring amplitude noise. LD: pump laser diode, ZnSe: 5 mm AR-coated substrate for pulse re-compression, L1: parabolic silver mirror (f=7 mm), PPLN: periodically-poled lithium niobate for second-harmonic generation, L2: BK-7 lens (f=15 mm), BP: bandpass filter at 1.2 μm (*Thorlabs*, FB1200-10), ND: neutral density filter wheel, M1: dielectric mirror (*Thorlabs*, BB1-E03), L3: AR-coated UVFS lens (f=50 mm), PD: biased InGaAs photodetector, RF analyzer: radio-frequency spectrum analyzer for noise detection. Non-labelled mirrors are coated for high reflection.